\documentclass[aps,prb,twocolumn,showpacs,superscriptaddress]{revtex4-1}

\usepackage{graphicx} % Include figure files

\begin{document}

\title{Tuning of the electronic and optical properties of single layer
black phosphorus by strain}

\author{Deniz \c{C}ak{\i}r}
\email{deniz.cakir@uantwerpen.be} \affiliation{Department of
Physics, University of Antwerp, Groenenborgerlaan 171, 2020 Antwerpen, Belgium}

\author{Hasan Sahin}
\email{hasan.sahin@uantwerpen.be} \affiliation{Department of Physics,
University of Antwerp, Groenenborgerlaan 171, 2020 Antwerpen, Belgium}

\author{Fran\c{c}ois M. Peeters}
\email{francois.peeters@uantwerpen.be} \affiliation{Department of
Physics, University of Antwerp, Groenenborgerlaan 171, 2020 Antwerpen, Belgium}

\date{\today}

\begin{abstract}

Using first principles calculations we showed that the electronic and optical properties of single layer 
black phosphorus (BP) depend strongly on the applied strain. Due to the strong 
anisotropic atomic structure of BP, its electronic conductivity and optical 
response are sensitive to the magnitude and the orientation of the applied 
strain.  We found that the inclusion of many body effects is essential for the correct
description of the electronic properties of monolayer BP; for example while the 
electronic gap of strainless BP is found to be 0.90 eV by using semilocal 
functionals, it becomes 2.31 eV when many-body effects are taken into account 
within the G$_0$W$_0$ scheme. Applied tensile strain was shown to significantly  
enhances electron transport along zigzag direction of BP. Furthermore, biaxial strain is able to 
tune the optical band gap of monolayer BP from 0.38 eV (at -8\% strain) to 2.07 eV (at 5.5\%). 
The exciton binding energy is also sensitive to the magnitude of the applied strain. 
It is found to be 0.40 eV for compressive biaxial 
strain of -8\%, and it becomes 0.83 eV for tensile strain of 4\%. Our 
calculations demonstrate that the optical response of BP can be significantly tuned using 
strain engineering which appears as a promising way to design novel photovoltaic 
devices that capture a broad range of solar spectrum. 
\end{abstract}


\maketitle

\section{Introduction}

With the synthesis of single graphene\cite{gr1,gr2,gr3} layers, a new era of
two-dimensional (2D) monolayer materials" emerged in condensed matter physics. It is
believed that further advances in synthesis and fabrication of those monolayer
crystals will pave the way in the exploration of many novel materials with 
exotic functionalities. Nowadays, single layers of
functionalized graphenes\cite{ch1,ch2,cf1,cf2,cf3,ccl,acsnano-hasan}, transition metal
dichalcogenides (TMDs),\cite{tmd1,tmd2,tmd3,tmd4,tmd5,tmd6,tmd7,tmd8} boron
nitride (BN)\cite{bn1} are readily accessible and some applications in 
nanoscale devices have been demonstrated. 
Although graphene is a fascinating 2D material, the lack of a bandgap in
its electronic spectrum leads to a search for similar ultra-thin
materials having a non-zero band gap. 

Recently, the successful synthesis of single layer phosphorus crystal
"phosphorene"  (so called black phosphorus (BP)) triggered interest in
this material.\cite{p1,p2,p4,chem-rev}  Single layer BP is an appealing material that can
be implemented into various electronic device applications including gas sensor~\cite{sensor}, p-n
junction~\cite{pn}, solar cell application~\cite{solar-cell} and field 
effect transistor (FET) due to its sizable band gap ($\sim$ 0.9 eV) and
higher carrier mobility as compared
to MoS$_2$\cite{p1,p2,bus,high-mob,xia,BP-characater}. Li et al.\cite{p1}
fabricated FETs based on few-layer black phosphorus crystals
and achieved reliable transistor performance at room temperature. Moreover, the 
stability and anisotropic structural properties of monolayer phosphorene were 
predicted by Liu et al.\cite{p2}. They observed a high on-current, 
a high hole field-effect mobility and a high on/off ratio in few-layer phosphorene FETs. 
Buscema et al.\cite{bus} demonstrated that black phosphorus is an
appealing candidate for tunable photodetection applications due its unique
properties such as (i) FETs allowing for ambipolar operation in the dark state
and (ii) broadband (from the visible region up to 940 nm) with fast detection
(rise time of about 1 ms) when illuminated. In a recent theoretical study by
Dai et al.\cite{dai} it was shown that the direct bandgap of phosphorene 
depends on the number of layer(from 0.3 to 1.5 eV) and that a 
vertical electric field can be used to tune this bandgap. 

Recently, accurate tight-binding (and GW) description of the electronic band 
structure of one to four layers black phosphorus and the importance of the 
interlayer hoppings were reported.\cite{rud}  In addition,
phosphorene nanoribbons (PNRs) were investigated using density functional
theory\cite{tran,han-ribbon}. Tran et al.\cite{tran} reported the electronic
structure and optical absorption spectra of monolayer phosphorene nanoribbons.
They showed that the band gaps of armchair ribbons scale as 1/$W^{2}$,
while zigzag PNRs exhibit a 1/$W$ behavior, where $W$ is the width of the 
nanoribbon. The direction-dependent width dependence of the band gap 
was attributed to the nonrelativistic behavior of electrons and holes along the 
zigzag direction and the relativistic behavior along the armchair direction. 
The respective PNRs' host electrons and holes with markedly different 
effective masses and optical absorption spectra were found to be suitable for a wide range 
of applications. Han et al.\cite{han-ribbon} showed that electronic properties 
(i.e. effective mass of charge carriers) of passivated phosphorene 
nanoribbons exhibit a strong dependence on the orientation and strain.

Although the electronic and structural properties of mono- and few-layer
phosphorenes have been investigated previously, the role of the strain on its
properties have remained an open question so far. 
Recently, BP was shown to have a negative Poisson's ratio\cite{poisson} and 
superior mechanical flexibility\cite{flex} 
that allows us to use it under extreme mechanical conditions. 
By using density functional theory and tight-binding models Rodin \textit{et 
al}.\cite{rodin} showed that the deformation in the direction normal to the 
crystal plane can be used to change the size of the energy bandgap and induce a 
semiconductor-metal transition. In addition Fei \textit{et al}.\cite{fei} reported that the
anisotropic free-carrier mobility of phosphorene can be controlled by applying
biaxial and uniaxial strain. In the present study we investigate how its 
electronic and optical properties change under biaxial  strain and calculate 
the exciton binding energy. We organize the paper as follows: electronic and 
transport properties of phosphorene under strain is presented in Sec. II and  
the optical response of monolayer phosphorene in the presence of biaxial strain is
investigated in Sec. III. Our results are summarized in Sec. IV.

\section{Transport properties}

\begin{figure}
\includegraphics[width=8.3cm]{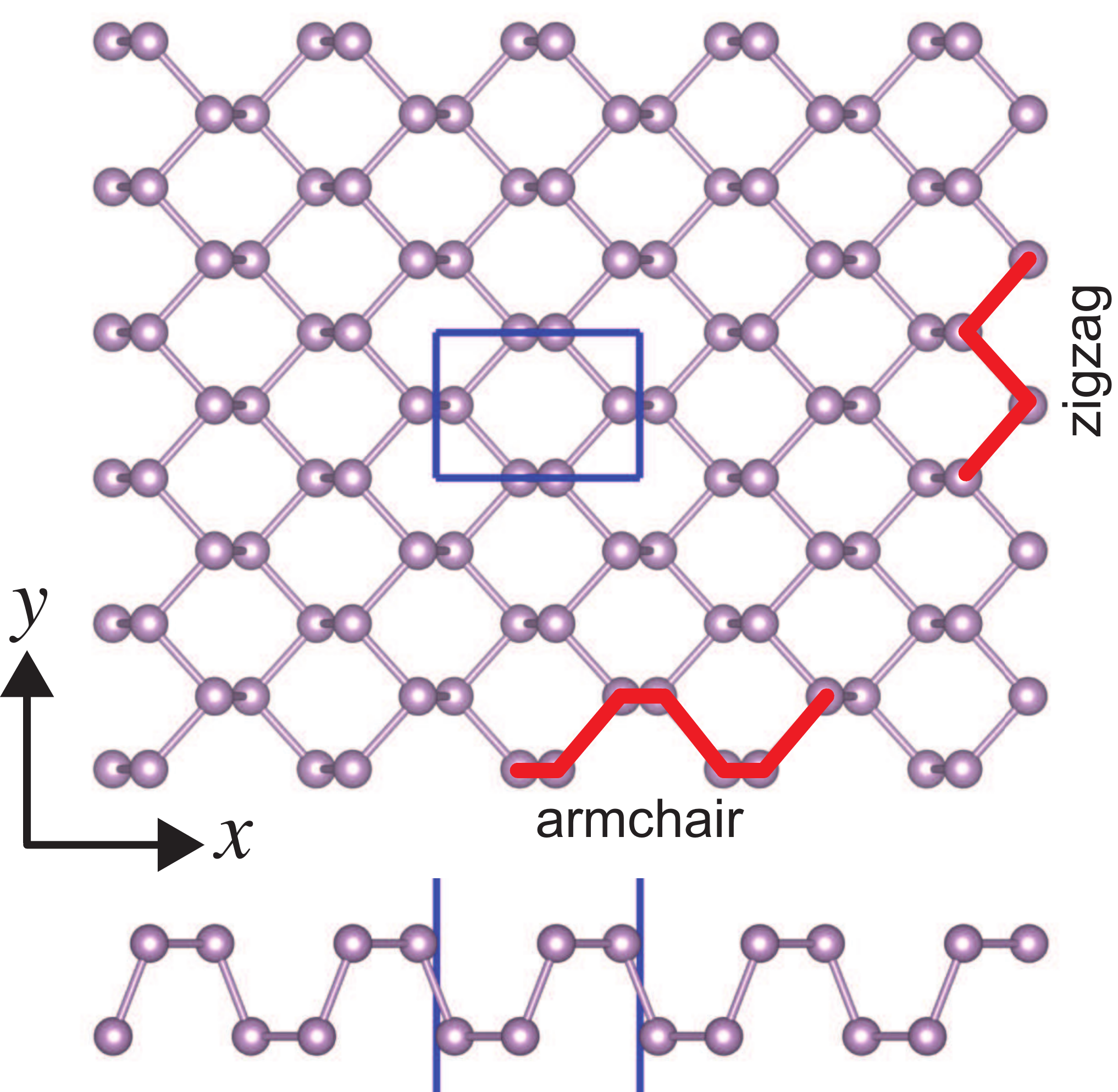}
\caption{\label{str} (Color online) Top and side views of
black phosphorus (BP) monolayer. The  unitcell of BP is shown by the blue 
rectangle.}
\end{figure}

Electronic transport through single layer phosphorene is calculated
by using the self-consistent nonequilibrium Green's functions (NEGF) technique
as implemented in TranSIESTA~\cite{transiesta}, which is interfaced with the
SIESTA code~\cite{siesta}. A double-$\zeta$ (plus polarization) numerical
orbital basis set is used for the P atom. We employ a  Troullier-Martins
norm-conserving pseudopotential~\cite{tm}, the GGA/PBE functional~\cite{PBE},
and an energy cutoff for the real-space mesh of 200 Ry. The electron transport 
is calculated along the armchair
(\textit{x}) and zigzag (\textit{y}) directions (see Fig.~\ref{str}). In order 
to get accurate transmission spectrum, the 2D Brillouin zone normal to the transport direction is
sampled by meshes composed of 100 \textbf{k}-points in the periodic direction.

\begin{figure}
\includegraphics[width=8.3cm]{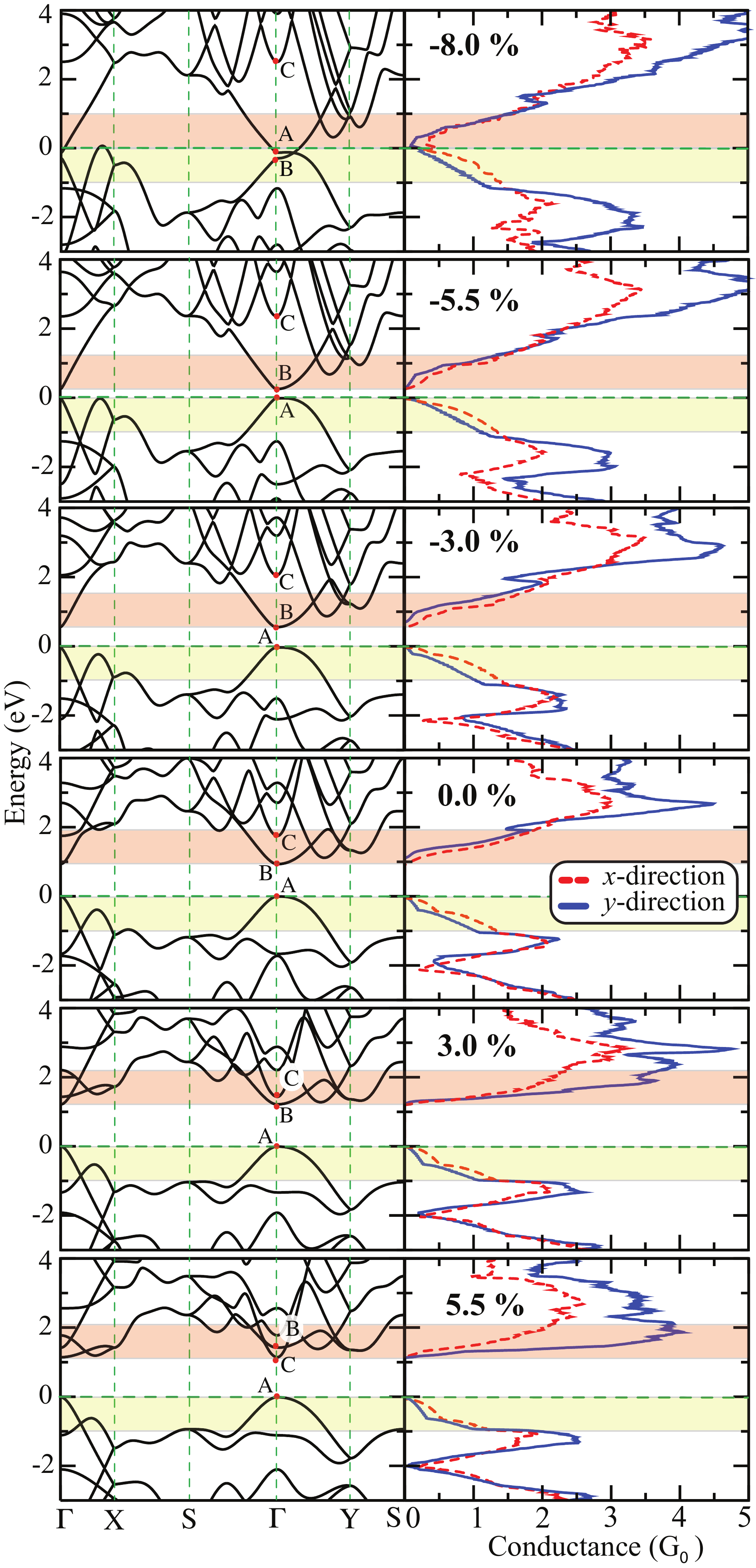}
\caption{\label{transport} (Color online) Band structure and conductance for 
single layer BP for different strain values. A, B, and
C mark the top of the valence band, bottom of the conduction band, and the second lowest
conduction band at the  $\Gamma$ point, respectively.  Shaded regions depict 
the upper part of the valence and lower part of the conduction bands. Here 
the quantum of conductance is G$_{0}$ = $\frac{2e^{2}}{h}$.}
\end{figure}

\begin{figure}
\includegraphics[width=8.3cm]{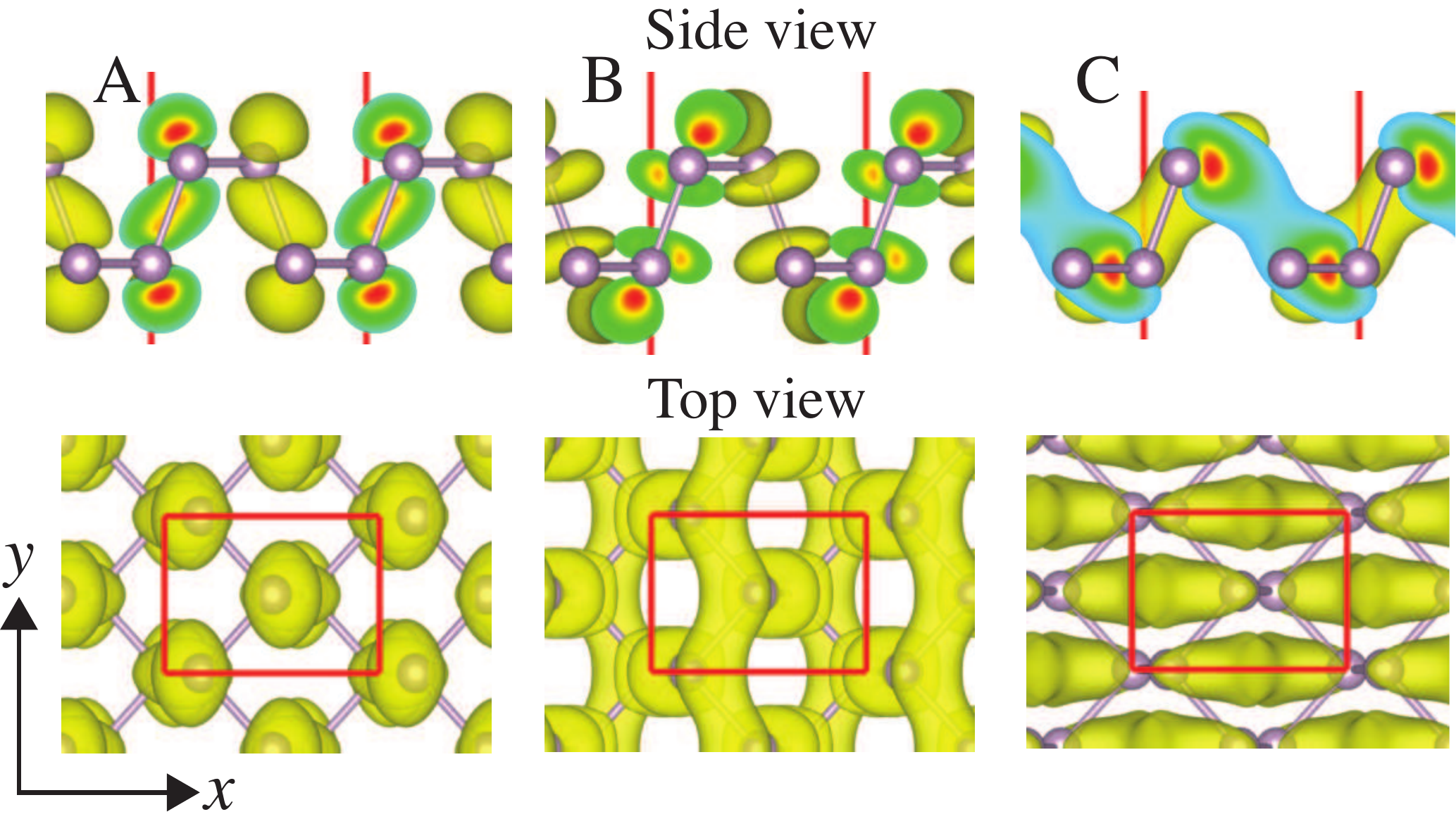}
\caption{\label{chg-den} (Color online) 3D band decomposed charge densities of
band edges which are labelled as A, B and
C  in Fig.~\ref{transport} for strainless BP.}
\end{figure}

Fig.~\ref{transport} shows the evolution of the band structure of the 
single layer BP with the applied in-plane biaxial tensile and compressive 
strain ($\varepsilon_{xy}$).  Strain-free single layer BP is a direct 
band gap semiconductor with a GGA/PBE band gap of 0.9 eV at the $\Gamma$ point. 
It is noteworthy to mention that BP has a  highly anisotropic band dispersion
around the band gap.  In other words, the top of the valence band and the bottom 
of the conduction band have a much larger  dispersion along the $\Gamma$-X  
direction as compared to the rather flat bands along the $\Gamma$-Y direction, 
resulting in significantly anistropic electronic properties. For instance, it 
was recently shown that the effective mass of electrons and holes are  highly 
anisotropic.~\cite{fei} For a better insight, in Fig.~\ref{chg-den},  the 
real-space wave functions corresponding to the top of the valence band  (marked 
as A),  bottom of the  conduction band (B) 
and the second lowest energy conduction band (C) at the $\Gamma$ point are 
depicted for strainless BP.  As seen in Fig.~\ref{chg-den},  
these three bands display very different spatial characters. 
The valence band edge (point A) has a non-bonding character in the $y$ and an anti-bonding 
character in the $x$ direction.  The conduction band edge (point B), which is 
dominated by the $p_z$-orbital, exhibits a bonding character along the $y$ 
direction. In contrast, it has a non-bonding nature in the $x$ direction.

For strain values of -5.5\% $\leq$ $\varepsilon_{xy}$ $\leq$ +5.5\%, BP monolayer remains
a direct band gap semiconductor at the $\Gamma$ point. At 
$\varepsilon_{xy}$=-5.5\%, direct and indirect band gap values are very close. 
Although not shown in Fig.~\ref{transport},
when strain is -6\%, BP becomes an indirect band gap semiconductor and a
semiconductor-to-metal transition is observed above -6.5\%. Interestingly, for 
uniaxial strains (either along zigzag or armchair directions), 
direct-to-indirect and semiconductor-to-metal transitions occur at much 
larger strain values.\cite{bp-strain-2} The band gap of BP exhibits a strong 
strain dependence. While compressive biaxial strain gives rise to a band gap 
lowering, tensile strain opens up the band gap up to 
$\varepsilon_{xy}$=4\%, but then reduces it at $\varepsilon_{xy}$=5.5\%. 
The reason for lowering in the band gap under larger tensile 
strain is that while the band edge B moves away from the band edge A, 
the edge C has the opposite tendency. Up to a critical tensile strain value 
at which the energy of the band edge C becomes lower than that of the band edge 
B, the band gap increases with strain. Once tensile strain exceeds the critical 
value, the band gap starts to decrease. For instance, while the band gap (at 
GGA-PBE level) is found to be ~0.25 eV at $\varepsilon_{xy}$=-5.5\%, it becomes 
1.19 eV at $\varepsilon_{xy}$=4\% and drops down to 1.03 eV 
at $\varepsilon_{xy}$=5.5\%.  As seen in Fig.~\ref{transport}, 
at $\varepsilon_{xy}$=-8\%,  A (B) point moves above (below) the 
Fermi level and BP is a metal for this strain value. 
   
One would expect that the observed variation in the electronic structure 
under strain can translate into a modulation in the electron conduction.
The right column of Fig.~\ref{transport} shows the conductance of BP monolayer as a
function of strain. Consistent with the band structure plots, the band gap 
values (within -5.5\% $\leq$ $\varepsilon_{xy}$ $\leq$ 4\%) calculated in 
transport calculations decreases (increases) with compressive (tensile) 
strain. Due to the highly anisotropic electronic properties, conductance of monolayer BP
shown in Fig.~\ref{transport} exhibits a strong direction and strain dependence. 
In contrast, for MoS$_2$ monolayer, it was found that the electrical 
conductivity does not depend on the direction of the current
in the presence of isotropic strain\cite{heine}. Applying
external strain switches the band ordering at the $\Gamma$-point of
the conduction band, see Fig.~\ref{transport}. 

As denoted in 
Fig.~\ref{chg-den}, since the band edge B has a bonding (nonbonding) character 
along the $y$($x$)-direction, the expansion of BP monolayer (especially along 
$y$ direction) lowers the binding energy of this band, leading to an upward 
shift. In contrast, the band edge C has an antibonding character along the 
$x$-direction, and thus applying tensile strain results in a downward 
shift. However, compressive strain shifts the band edge C towards the higher 
energies to prevent the population of this band. For tensile strain, the 
conductivity along the \textit{y} direction around the conduction band edge 
significantly enhances as a result of strain induced rotation of the electronic
conduction around the $\Gamma$-point. This can be clearly seen in 
Fig.~\ref{transport} when comparing the evolution of the conductance with 
strain within the shaded region in the conduction band. The second lowest empty 
band (which is labelled as C in Fig.~\ref{transport}) moves down with 
increasing strain, whose effect can be directly seen in the electrical 
conduction along the \textit{y} direction.  At $\varepsilon_{xy}$=5.5\%, this 
band has lower energy than band B which is the lowest energy empty band for 
-5.5\% $\leq$ $\varepsilon_{xy}$ $<$ 4\%. In the vicinity of band edge C, the 
effective mass of the charge carriers in the \textit{y} direction
($\Gamma$-Y) is much smaller than that in the \textit{x} direction ($\Gamma$-X)\cite{fei}.
In contrast, the effective mass of the charge carriers in the conduction band 
minimum (B point) exhibits the opposite behavior.

\begin{figure*}
\includegraphics[width=15.0cm]{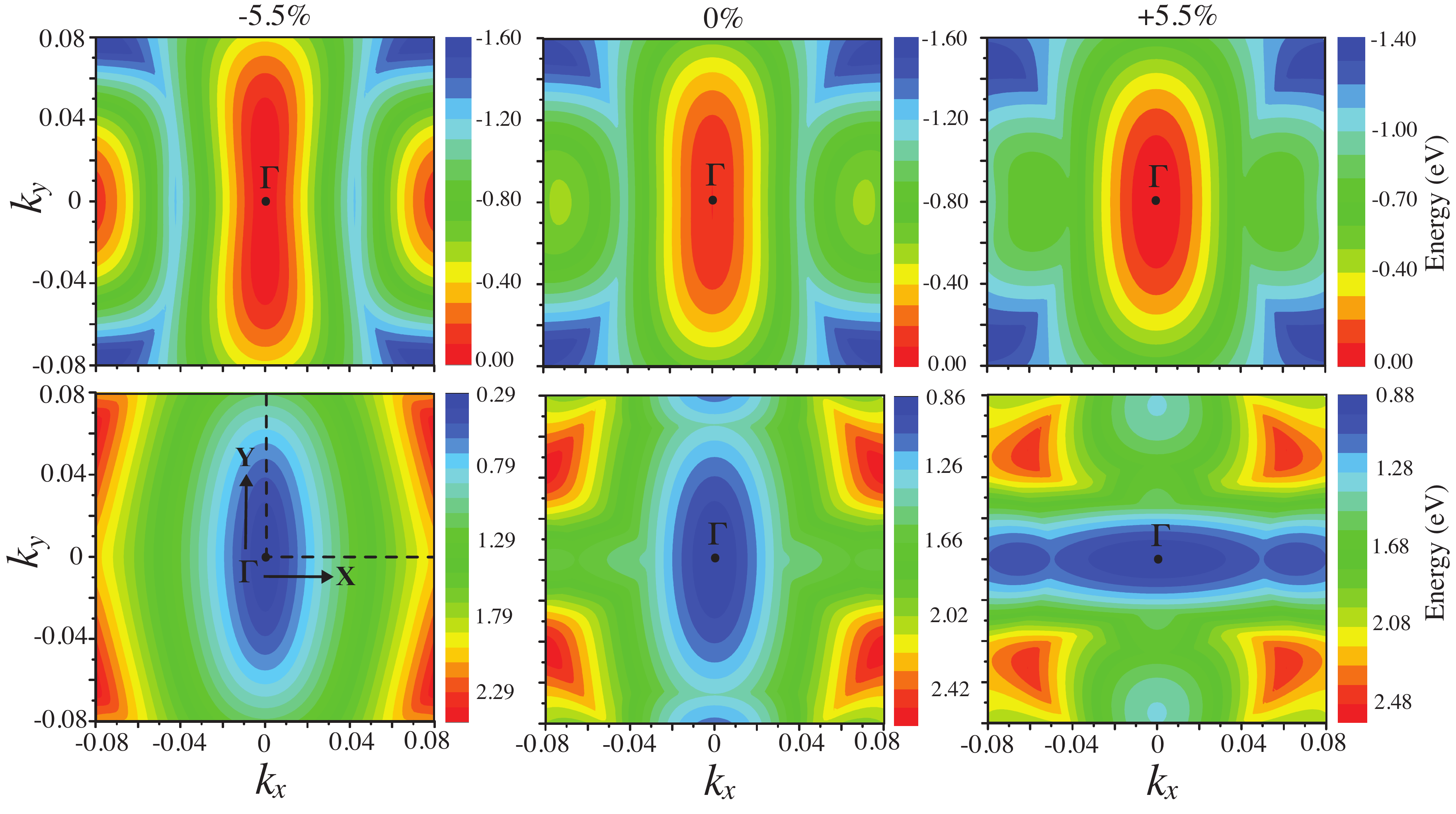}
\caption{\label{contour} (Color online) Contour plots of the valence (top) and 
the conduction (bottom) bands around the $\Gamma$-point for different strain 
values calculated with GGA-PBE functional. }
\end{figure*}

Fig.~\ref{contour} displays
contour plots of the valence and conduction bands around the $\Gamma$ point,
which illustrates the anisotropic nature of those bands. The change of band 
order at $\Gamma$  
can be easily followed in Fig.~\ref{contour}. 

Another important point is that 
the high value conductance peaks also move towards lower energies with 
increasing strain, which significantly increases the conductivity around
the conduction band edge. Especially, the effect of strain on the different 
bands are more pronounced around the $\Gamma$-point. As seen in 
Fig.~\ref{transport}, the valence band along the $\Gamma$-Y direction is flat, 
resulting in a very large effective mass for holes. Therefore, the conductivity 
is quite low for this band along the $\Gamma$-Y direction (i.e., zigzag 
direction), see Fig.~\ref{transport}. In addition, the valence band is found to 
be insensitive to the applied strain, which results in a quite similar 
conductivity for all strain values considered in this study. It appears 
that tensile strain dramatically modifies the electron transport properties of 
single layer BP.

\section{Optical properties}

One could expect that the excitonic effects are 
influential on the optical properties of BP due to 
the weak screening and reduced dimensionality. 
For a correct description of the optical properties of BP, many body interactions
(electron-electron and electron-hole interaction for instance) must be taken
into account. In this work, we calculated the optical spectra of monolayer BP
under both compressive and tensile biaxial strains ($\varepsilon_{xy}$)
using the Bethe-Salpeter-equation (BSE) method\cite{bse-1,bse-2} within the 
Vienna ab-initio simulation package (VASP)\cite{VASP,VASP1}. First, hybrid-DFT 
calculations were performed within the HSE06\cite{hse-1,hse-2,hse-3} approach 
for single layer BP structure that was optimized using GGA-PBE. This is followed 
by one-shot GW (i.e., G$_0$W$_0$) calculations to obtain the quasiparticle
excitations\cite{gw-1,gw-2,gw-3,gw-4,gw-5,gw-6}. Finally, we carried out BSE
calculations on top of G$_0$W$_0$ in order to obtain the optical adsorption spectra by
including excitonic effects using the Tamm-Dancoff 
approximation\cite{Tamm-Dancoff}.

The BSE calculations were performed on a 9$\times$13$\times$1 \emph{k}-mesh 
within the Monkhorst-Pack scheme.\cite{monk} The energy cutoff for the 
wavefunctions and for the response functions were set to 400 eV and 200 eV, 
respectively. Since the number of empty bands significantly influences the 
relative position of the quasiparticle energy states, we tested the convergence 
for 112 and 326 empty bands. The calculated quasiparticle gap and exciton 
binding energy are then converged within 0.05 eV. The six highest occupied 
valence bands and six lowest unoccupied conduction bands were included as basis 
for the excitonic states. Since GW calculations require a sufficiently large 
vacuum region, we use a vacuum region of at least 15 \AA~ to avoid spurious 
interaction between the periodic images. A complex shift of $\eta$=0.05 eV was 
employed to broaden the calculated absorption spectra.

\begin{table}
\caption{\label{table1} Calculated electronic gap ($E_{gap}$), optical gap 
($E_{opt}$) and exciton binding energy ($E_{exc}$) for different biaxial strain 
($\varepsilon_{xy}$) values in units of eV. Here, M stands for metallic 
behavior. Since BP is an indirect band gap semiconductor for the strain values 
of -6\% and -8\%, we give the indirect band gap value. The direct gap is 0.02 
eV bigger than the indirect one.}
\begin{tabular}{lccccccccc}
\hline\hline
           & $\varepsilon_{xy}= $ & --8\% & --6\% &--4\%  &  0   &  +4\%  & +5.5\% \\
\hline
  $E_{gap}^{PBE}$    &  &   M  & 0.26 &  0.47 & 0.90 & 1.19 & 1.03  \\
  $E_{gap}^{HSE06}$  &  & 0.32 & 0.85 &  1.09 & 1.59 & 1.95 & 1.80  \\
  $E_{gap}^{G_0W_0}$ &  & 0.78 & 1.28 &  1.69 & 2.31 & 2.87 & 2.72  \\
  $E_{opt}$          &  & 0.38 & 0.75 &  1.07 & 1.61 & 2.04 & 2.07  \\
  $E_{exc}$          &  & 0.40 & 0.53 &  0.62 & 0.70 & 0.83 & 0.65  \\
  \hline \hline
\end{tabular}
\end{table}

\begin{figure}
\includegraphics[width=9cm]{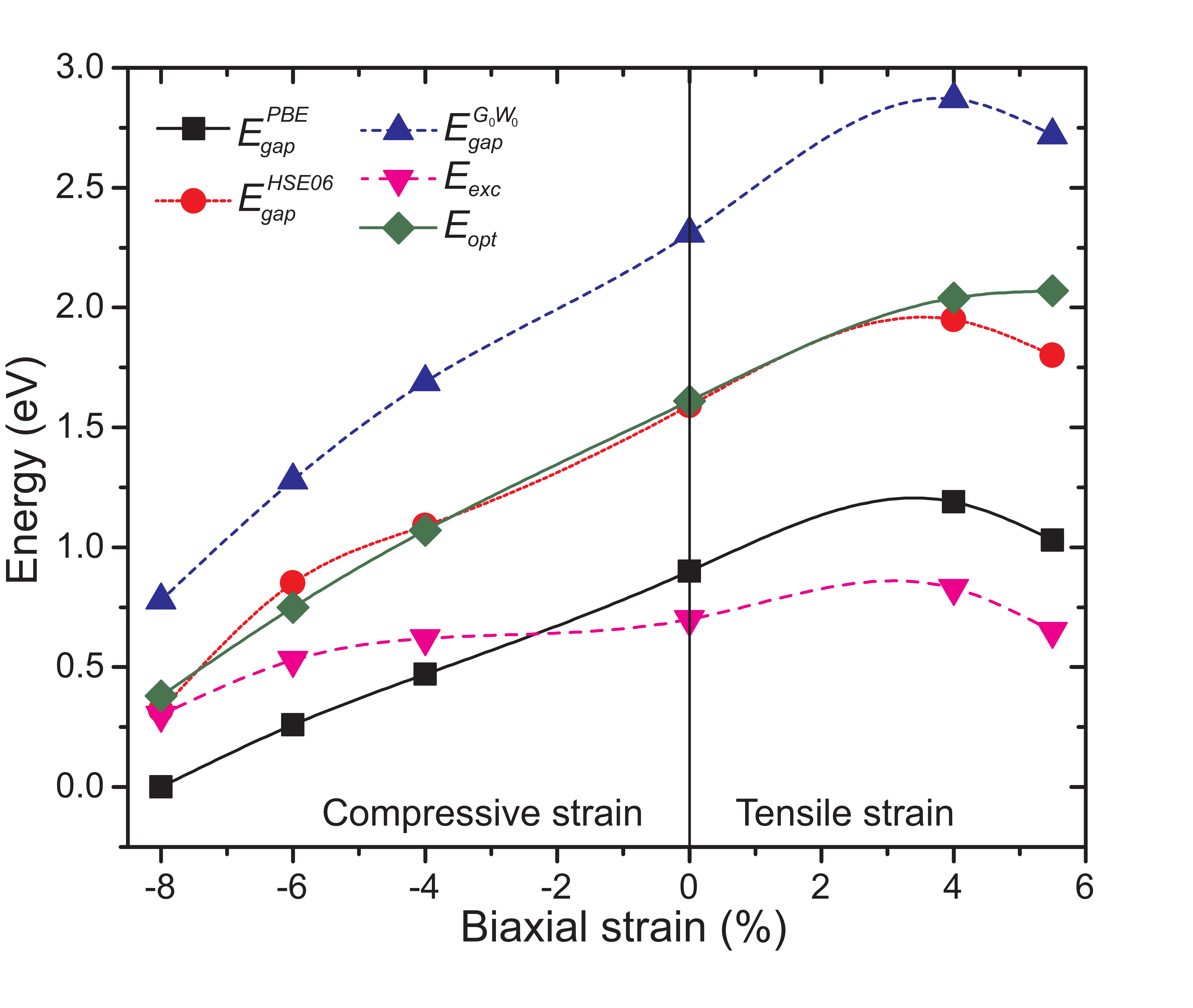}
\caption{\label{plot} (Color online)  Electronic band gap ($E_{gap}$),
optical gap ($E_{opt}$) and exciton binding energy ($E_{exc}$)
as a function of biaxial strain for different exchange-correlation functionals.}
\end{figure}

Table~\ref{table1} summarizes the calculated electronic gap ($E_{gap}$), 
optical gap ($E_{opt}$) and exciton binding energy ($E_{exc}$ = 
$E^{G_0W_0}_{gap}$ - $E_{opt}$) for different strain values. The variation 
of the  electronic and the optical properties of BP as a function of biaxial 
strain is plotted in Fig.~\ref{plot}. As seen in Table~\ref{table1} and Fig.~\ref{plot}, 
using hybrid functionals significantly enlarges $E_{gap}$. While $E_{gap}$ for 
strainless BP monolayer is calculated as 0.90 eV, in consistent with previous 
calculations, in GGA-PBE, it becomes 1.59 eV when using HSE06. 
Inclusion of many body effects ($G_0W_0$) further increases $E_{gap}$ to 2.31 
eV (see Table~\ref{table1}). Previous theoretical studies predicted that 
$E_{gap}$ of single layer BP increases (decreases) under tensile (compressive) 
strain~\cite{bp-strain-2}. We observe that not only $E_{gap}$ but also $E_{opt}$ 
and $E_{exc}$ are very sensitive to the applied strain. Therefore, by tuning 
strain, optical properties of BP can be easily modified. Adversely, $E_{exc}$ 
hardly changes under strain for MoS$_2$ monolayer.\cite{yakobson} The 
experimental optical gap\cite{p2} for single layer BP was shown to be around 
1.45 eV, in fair agreement with the present calculated value of 1.61 eV. 
In a recent experiment, since BP is placed on a SiO$_2$ surface, it 
is expected that screening effects weakens the binding and decreases
the exciton binding energy, in qualitative agreement with our simulations\cite{p2}.
Due to the reduced dimensionality  and weak screening, our calculations predict a 
large exciton binding energy for strainless monolayer BP of 0.7 eV, which is in 
good agreement with recent theoretical works\cite{bp-ex,bp-ex-2}.
Note that the calculated exciton binding energy for BP is comparable to those 
found for other monolayer semiconductors such as MoS$_2$. $E_{exc}$
has been found to be in the order of 1 eV for single layer
MoS$_2$~\cite{Louie,yakobson,AVK-1,AVK-2,mose-ex}. Another point is that 
tensile strain up to $\varepsilon_{xy}$=4\%  enhances the exciton binding due 
to the weakening of the dielectric screening as a result of a significant 
increase of the electronic band gap. It is worth to mention that when one uses 
PBE wavefunctions and eigenvalues as inputs for the GW calculations instead of 
HSE06, $E_{opt}$ is found to be 1.30 eV instead of 1.61 eV, pointing out
to the significance of the HSE step in calculating the
quasiparticle and optical gap. 

\begin{figure*}
\includegraphics[width=17cm]{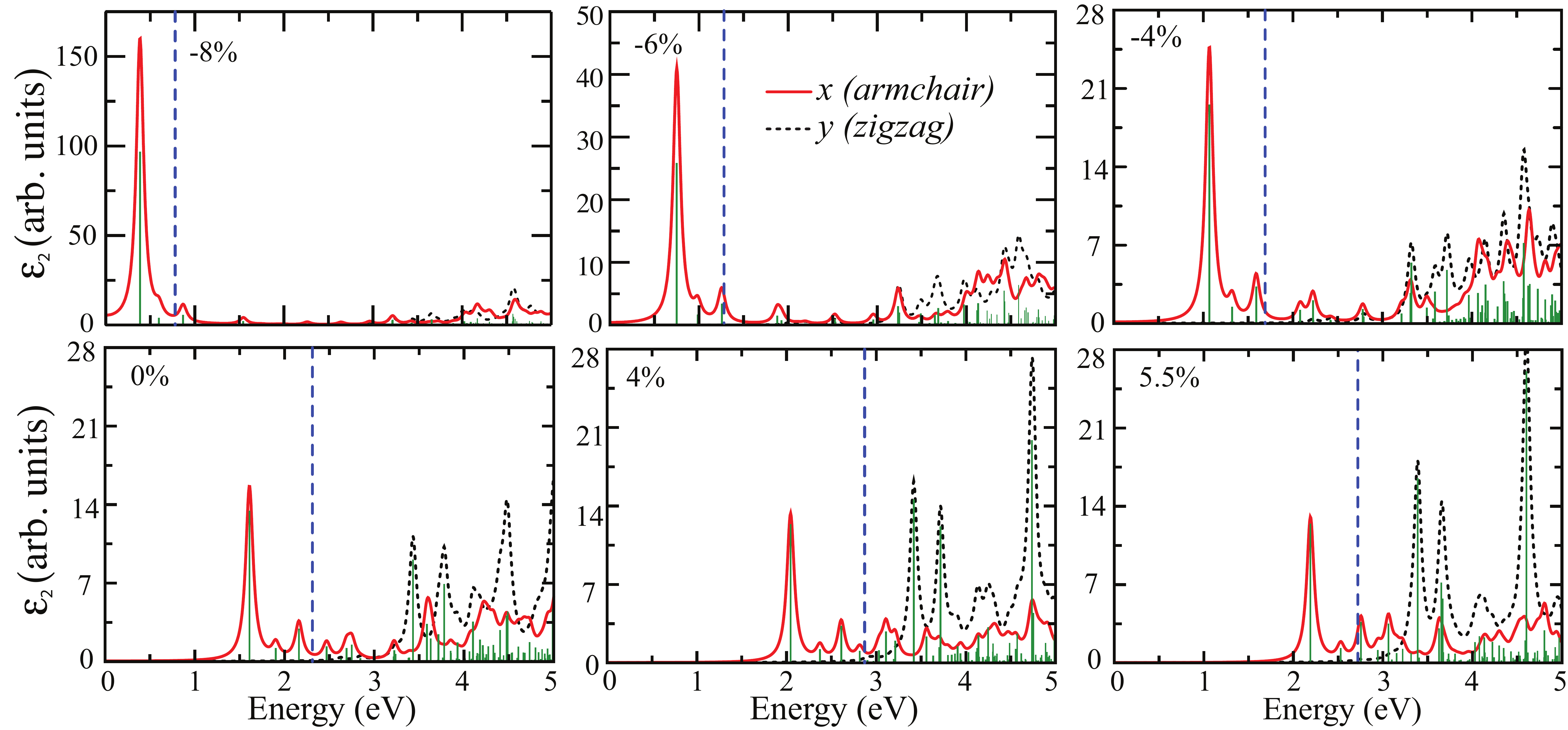}
\caption{\label{optical} (Color online) G$_0$W$_0$+BSE absorption spectra for
monolayer BP as a function of strain. Blue vertical dashed lines mark the 
electronic
band gap calculated at the level of G$_0$W$_0$. Green vertical lines represent
the relative oscillator strengths for the optical transitions.}
\end{figure*}

As mentioned in the previous section, GGA-PBE functional predicts that single 
layer BP undergoes a semiconductor-to-metal transition when 
$\varepsilon_{xy}$=-8\%. However, both HSE06 and G$_0$W$_0$ predict that BP is 
an indirect band gap semiconductor at this 
strain value with a gap of 0.32 (0.78) eV with HSE06 
(G$_0$W$_0$). In line with the present study, it was previously shown that 
the calculated vertical compressive strain value that is needed to induce a 
semiconductor-to-metal transition in bilayer MoS$_2$ was found to be much 
larger in GW than that in GGA-PBE\cite{gw-strain-mos2}. Therefore, to 
predict the transition 
point accurately, GW calculations are essential.

Fig.~\ref{optical} displays the optical absorption spectra ($\varepsilon_2$) of 
single layer BP for light polarized along \textit{x} (armchair) and \textit{y} 
(zigzag) directions for different strain values. We also show the electronic 
band gap values calculated with $G_0W_0$. Similar to the transport properties, 
the optical absorption spectra displays a strong orientation dependence.
It is clear that excitonic effects largely affect the optical spectra of BP. 
For all strain values, we notice an absorption peak 
along the \textit{x} direction, since BP absorbs $x$-polarized light due to its anisotropic
electronic structure. This peak is observed at 1.61 eV for strain-free BP, and it
moves to lower (higher) energy under compressive (tensile) strain. 
We find three bound exciton states when $\varepsilon_{xy}$=-4\% and  $\varepsilon_{xy}$=0 and four bound 
exciton states for $\varepsilon_{xy}$=4\% with wavefunction localized in the 
\textit{x} direction. Adversely, there is no bound exciton that is localized in the \textit{y} direction, 
meaning that BP is transparent to polarized light along the \textit{y} direction and  when an 
exciton is formed, it strongly localizes along the \textit{x} direction. 
Increasing tensile strain from 4\% to 5.5\% decreases the electronic band gap 
(at the G$_0$W$_0$ level ) and optical gap to 2.72 and 2.07 eV, respectively. 
Similarly, exciton binding weakens and  $E_{exc}$  for $\varepsilon_{xy}$=5.5\%  
becomes very close to that for $\varepsilon_{xy}$=-4\%. In addition, at 
$\varepsilon_{xy}$=5.5\%, the number of bound exciton states  drops from four to 
two. Interestingly, while $E_{opt}$ and  $E_{exc}$  increase up to 
$\varepsilon_{xy}$=4\% and then decrease at $\varepsilon_{xy}$=5.5\%, the first 
absorption peak always moves to higher  energies as the strain changes from 
compressive to tensile.

These results confirm that strain has a significant effect on the optical
properties. For instance, the absorption energy window of monolayer BP is 
tunable via strain. While BP monolayer under -4\% compressive strain absorbs infrared light (energies
between 1.1 eV and 1.7 eV),  strainless BP can absorb the low energy part of the 
visible light spectrum (up to 2.4 eV). In addition, applying 4\% tensile strain 
makes BP active almost over the whole visible range. Similar to interesting 
properties revealed by recent experiments on  MoS$_2$\cite{nature-phot},  
strain engineering can be used to tune the optical 
properties of photovoltaic devices made from a strain engineered BP
monolayer.

\section{Conclusion}

In conclusion, we performed first-principles calculations in order to determine the
effect of biaxial strain on the electronic transport and optical properties of 
BP. We find that the optical absorption spectrum and the carrier conductivity 
are highly anisotropic and strongly depend on the amount of applied strain. The 
large variation in the calculated electronic band gap of BP shows clearly the 
importance of many body effects in electronic structure calculations. The 
exciton binding  is found to be large due to the reduced dimensionality and 
weak screening.  $E_{exc}$ is 0.70 eV for stainless BP, and it increases 
(decreases) to 0.83  eV (0.40 eV) at the strain value of  +4\% (-8\%). In 
addition the optical gap varies by 1.5 eV when the strain changes from 
compressive to tensile. Here, strain engineering appears as a quite exciting 
way to tune the optical response and the electrical conductivity of BP 
which is potentially useful for device applications including flexible
electronics and optical devices due to its atomically thin crystal 
structure and direct band gap.   

\section{Acknowledgments}
This work was supported by the Flemish Science Foundation (FWO-Vl)
and the Methusalem foundation of the Flemish government.
Computational resources were provided by TUBITAK ULAKBIM,
High Performance and Grid Computing Center (TR-Grid e-Infrastructure),
and HPC infrastructure of the University of Antwerp (CalcUA) a
division of the Flemish Supercomputer Center (VSC), which is funded by
the Hercules foundation. H.S. is supported by a FWO Pegasus Marie
Curie-long Fellowship. D.C. is supported by a FWO Pegasus-short Marie 
Curie Fellowship.


\begin{thebibliography}{99}
%
 \bibitem{gr1} K. S. Novoselov, A. K. Geim, S. V. Morozov, D.
Jiang, Y. Zhang, S. V. 
 Dubonos, I. V. Grigorieva, and A. A. Firsov, Science
\textbf{306}, 666 (2004).
% 
 \bibitem{gr2} K. S. Novoselov, A. K. Geim, S. V. Morozov, D.
Jiang, M. I. Katsnelson,
 I. V. Grigorieva, S. V. Dubonos, and A. A. Firsov, Nature (London) 
\textbf{438}, 197 (2005).
% 
% 
 \bibitem{gr3} K. S. Novoselov, Z. Jiang, Y. Zhang, S. V. Morozov,
H. L. Stormer, U. Zeitler, 
 J. C. Maan, G. S. Boebinger, P. Kim, and A. K. Geim, 
 Science \textbf{315}, 1379 (2007).

\bibitem{ch1} J. O. Sofo, A. S. Chaudhari, and G. D. Barber,
Phys. Rev. B \textbf{75}, 153401 (2007).

\bibitem{ch2} H. Sahin, C. Ataca, and S. Ciraci, Appl. Phys. Lett. \textbf{95},
222510 (2009).
% 
 \bibitem{cf1} R. R. Nair, W. Ren, R. Jalil, I. Riaz, V. G.
Kravets, L. Britnell, P. Blake, F. Schedin, A. S.  Mayorov, 
 S. Yuan, M. I. Katsnelson, H. M. Cheng, W. Strupinski, L. G. Bulusheva, A. V.
Okotrub, I. V. Grigorieva, A. N. Grigorenko, K. S. Novoselov, 
 and A. K. Geim, Small \textbf{6}, 2877 (2010).
% 
 \bibitem{cf2} H. Sahin, M. Topsakal, and S. Ciraci, Phys. Rev. B
\textbf{83}, 115432 (2011).
%
 \bibitem{cf3} H. Peelaers, A. D. Hernandez-Nieves, O. Leenaerts, B.
Partoens, and F. M. Peeters, Appl. Phys. Lett. \textbf{98}, 051914 (2011).
% 
 \bibitem{ccl} H. Sahin and S. Ciraci, J. Phys. Chem. C
\textbf{116}, 24075 (2012).
% 

\bibitem{acsnano-hasan}
A. L. Walter, H. Sahin, K.-J. Jeon, A. Bostwick, S. Horzum, R. Koch, 
F. Speck, M. Ostler, P. Nagel, M. Merz , S. Schupler, L. Moreschini, 
Y. J. Chang, T. Seyller, F. M. Peeters, K. Horn, and E. Rotenberg, 
ACS Nano \textbf{8}, 7801 (2014).

\bibitem{tmd1} M. Chhowalla, H. S.  Shin, G. Eda, L. J. Li,
K. P. Loh, and H. Zhang, Nat. Chem. \textbf{5}, 263 (2013).
% 
 \bibitem{tmd2}  S. Tongay, H. Sahin, C. Ko, A. Luce, W. Fan, K. Liu, J.
Zhou, Y. S. Huang, C. H. Ho, J. Yan, 
 D. F. Ogletree, S. Aloni, J. Ji, S. Li, J. Li, F. M. Peeters, and J. Wu, 
 Nat. Commun. \textbf{5}, 3252 (2014).
% 
 \bibitem{tmd3} B. Sipos, A. F. Kusmartseva, A. Akrap, H. Berger, L.
Forro, and E. Tutis, Nat. Mater. \textbf{7}, 960 (2008).
% 
 \bibitem{tmd4} J. N. Coleman, M. Lotya, A. O'Neill, S. D. Bergin, P. J. King, 
U. Khan, K. Young, A. Gaucher, S. De, R. J. Smith, I. V. Shvets, S. K. Arora, 
J. J. Boland, J. J. Wang, J. F. Donegan, J. C. Grunlan, G. Moriarty, A.
Shmeliov, 
R. J. Nicholls, J. M. Perkins, E. M. Grieveson, K. Theuwissen, D. W. McComb, P.
D. Nellist, and V. Nicolosi, Science \textbf{331}, 568 (2011).
% 
 \bibitem{tmd5} A. Splendiani, L. Sun, Y. Zhang, T. Li, J.
Kim, C. Y. Chim, G. Galli, and F. Wang, Nano Lett. \textbf{10}, 1271 (2010).
%
 \bibitem{tmd6} Q. H. Wang, K. Kalantar-Zadeh, A. Kis, J. N. Coleman,
and M. S. Strano, Nature Nanotech. \textbf{7}, 699 (2012).
% 
 \bibitem{tmd7} H. Sahin, S. Tongay, S. Horzum, W. Fan, J. Zhou, J. Li, J.
Wu, and F. M. Peeters, Phys. Rev. B \textbf{87}, 165409 (2013).

\bibitem{tmd8}
D. \c{C}ak{\i}r, F. M. Peeters, and C. Sevik,
App. Phys. Lett. \textbf{104}, 203110 (2014).

\bibitem{bn1}K. K Kim, A. Hsu, X. Jia, S. M. Kim, Y. Shi, M. Hofmann, D. Nezich,
J. F. Rodriguez-Nieva, M. Dresselhaus, T. Palacios, and J. Kong, Nano Lett. 
\textbf{12}, 161 (2012).

%===============BP===================

\bibitem{p1} 
L. Li, Y. Yu, G. J. Ye, Q. Ge, X. Ou, H.
Wu, D. Feng, X. H. Chen, and Y. Zhang, Nat. Nanotechnol. \textbf{9}, 372 (2014).

\bibitem{p2} 
H. Liu, A. T. Neal, Z. Zhu, Z. Luo, X. Xu,
D. Tomanek, and P. D. Ye, ACS Nano \textbf{8}, 4033 (2014).

\bibitem{p4} 
E. S. Reich, Nature (London) \textbf{506}, 19 (2014).

\bibitem{chem-rev}
H. Liu, Y. Du, Y. Deng and P. D. Ye, Chem. Soc. Rev.,
DOI: 10.1039/c4cs00257a

\bibitem{sensor}
L. Kou, T. Frauenheim, and C. Chen,
J. Phys. Chem. Lett. \textbf{5}, 2675 (2014).

\bibitem{pn}
M. Buscema, D. J. Groenendijk, G. A. Steele, H. S.J. van der Zant and A. 
Castellanos-Gomez, Nat. Commun. \textbf{5}, 4651 (2014).

\bibitem{solar-cell}
J. Dai and X. C. Zeng, J. Phys. Chem. Lett. \textbf{5}, 1289 (2014).

\bibitem{bus} 
M. Buscema, D. J. Groenendijk , S. I. Blanter , G. A. Steele,
H. S. J. van der Zant, and A. Castellanos-Gomez, Nano Lett. \textbf{14} (6),
3347 (2014).

\bibitem{high-mob}
J. Qiao, X. Kong, Z.-X. Hu, F. Yang, and W. Ji,
Nat. Commun. \textbf{5}, 4475 (2014).

\bibitem{xia}
F. Xia, Han Wang, and Y. Jia, 
Nat. Commun. \textbf{5}, 4458 (2014).

\bibitem{BP-characater}
A. Castellanos-Gomez, L. Vicarelli, E. Prada, J. O. Island, 
K. L. Narasimha-Acharya, S. I. Blanter, D. J. Groenendijk, 
M. Buscema, G. A. Steele, J. V. Alvarez, H. W. Zandbergen, J. J. Palacios and 
H. S.J. van der Zant, 2D Mater. \textbf{1}, 025001 (2014).

\bibitem{dai} 
J. Dai and X. C. Zeng, J. Phys. Chem. Lett. \textbf{5}, 1289 (2014).

\bibitem{rud} 
A. N. Rudenko and M. I. Katsnelson, Phys. Rev. B \textbf{89},
201408(R) (2014).

\bibitem{tran}  
V. Tran and L. Yang, Phys. Rev. B \textbf{89}, 245407 (2014).

\bibitem{han-ribbon}
X. Han, H. M. Stewart, S. A. Shevlin, C.  A. Catlow, and Z. X. Guo,
Nano Lett., \textbf{14} (8), 4607 (2014).

\bibitem{poisson}
J.-W. Jiang	and H. S. Park,
Nat. Commun. \textbf{5}, 4427 (2014).

\bibitem{flex}
Q. Wei, and X. Peng, Appl. Phys. Lett. \textbf{104}, 251915 (2014). 

\bibitem{rodin} 
A. S. Rodin, A. Carvalho, and A. H. C. Neto, Phys. Rev.
Lett. \textbf{112}, 176801 (2014).

\bibitem{fei} 
R. Fei and L. Yang, Nano Lett. \textbf{14}, 2884 (2014).


%%%%%%%%%THEORY-transport%%%%%%%%%%%%%%%%%%


\bibitem{transiesta}
M. Brandbyge, J.-L. Mozos, P. Ordej\'{o}n, J. Taylor, and K.
Stokbro, Phys. Rev. B \textbf{65}, 165401 (2002).

\bibitem{siesta}
J. M. Soler, E. Artacho, J. D. Gale, A. Garcia, J. Junquera, P.
Ordej\'{o}n, and D. S\'{a}nchez-Portal, J. Phys.: Cond. Matter.
\textbf{14}, 2745 (2002).

\bibitem{tm}
N. Troullier and J. L. Martins, Phys. Rev. B \textbf{43}, 1993
(1991).

\bibitem{PBE}
J. P. Perdew, K. Burke, and M. Ernzerhof, Phys. Rev. Lett. \textbf{77}, 3865
(1996).

\bibitem{bp-strain-2}
X. Peng, Q. Wei, and A. Copple,
Phys. Rev. B \textbf{90}, 085402 (2014).

\bibitem{heine}
M. Ghorbani-Asl, S. Borini, A. Kuc, and T. Heine, Phys. Rev. B
\textbf{87}, 235434 (2013).


%%%%%%%%%%%%THEORY-OPTIC%%%%%%%%%%%%%

\bibitem{bse-1}
E. E. Salpeter and H. A. A Bethe, 
Phys. Rev. \textbf{84}, 1232 (1951).

\bibitem{bse-2}
G. Onida,  L. Reining, and A. Rubio, 
Rev. Mod. Phys. \textbf{74}, 601 (2002).

\bibitem{VASP} 
G. Kresse and J. Furthmuller,  Comput. Mater. Sci. \textbf{6}, 15 (1996).

\bibitem{VASP1}
G.Kresse and J.Furthm\"uller, Phys. Rev. B \textbf{54}, 11169 (1996).

\bibitem{hse-1}
J. Heyd, G. E. Scuseria, and M. Ernzerhof, J. Chem. Phys., \textbf{118},
8207 (2003).

\bibitem{hse-2}
J. Paier, M. Marsman, K. Hummer, G. Kresse, I. C. Gerber, and J. G.
\'{A}ngy\'{a}n, J. Chem. Phys. \textbf{125}, 249901 (2006).

\bibitem{hse-3}
J. Heyd, G. E. Scuseria, and M. Ernzerhof, J. Chem. Phys. \textbf{124},
219906 (2006).

\bibitem{gw-1}
L. Hedin, Phys. Rev. \textbf{139}, A796 (1965).

\bibitem{gw-2}
M. S. Hybertsen and S. G. Louie, Phys. Rev. B \textbf{34}, 5390 (1986).

\bibitem{gw-3}
R. W. Godby, M. Schlter, and L. J. Sham, Phys. Rev. B \textbf{37}, 10159 (1988).

\bibitem{gw-4}
M. Shishkin and G. Kresse, Phys. Rev. B \textbf{74}, 035101 (2006).

\bibitem{gw-5}
M. Shishkin and G. Kresse, Phys. Rev. \textbf{B} 75,
235102 (2007).

\bibitem{gw-6}
M. Shishkin, M. Marsman, and G. Kresse, Phys.  Rev. Lett. \textbf{99}, 246403 (2007).

\bibitem{Tamm-Dancoff}
J. Paier, M. Marsman, and G. Kresse, Phys. Rev. B \textbf{78}, 121201(R) (2008).

\bibitem{monk}
H. J. Monkhorst and J. D. Pack, Phys. Rev. B \textbf{13}, 5188 (1976).

%%%%%%%%%%%%%%%%%%%%%%bp

\bibitem{yakobson}
H. Shi, H. Pan, Y.-W. Zhang, and B. I. Yakobson,
Phys. Rev. B \textbf{87}, 155304 (2013).

 \bibitem{bp-ex}
V. Tran, R. Soklaski, Y. Liang, and L. Yang,
Phys. Rev. B \textbf{89}, 235319 (2014).

\bibitem{bp-ex-2}
A. S. Rodin, A. Carvalho, and A. H. Castro Neto,
Phys. Rev. B \textbf{86}, 075429 (2014).

\bibitem{Louie}
D. Y. Qiu, F. H. da Jornada, and S. G. Louie,
Phys. Rev. Lett. \textbf{111}, 216805 (2013).

\bibitem{AVK-1}
H. -P. Komsa and A. V. Krasheninnikov
Phys. Rev. B \textbf{86}, 241201(R) (2012).

\bibitem{AVK-2}
H. -P. Komsa and A. V. Krasheninnikov
Phys. Rev. B \textbf{88}, 085318 (2013).

\bibitem{mose-ex}
A. Ramasubramaniam, 
Phys. Rev. B \textbf{86}, 115409 (2012).

\bibitem{gw-strain-mos2}
S. Bhattacharyya and A. K. Singh,
Phys. Rev. B \textbf{86}, 075454 (2012).


\bibitem{nature-phot}
J. Feng, X. Qian, C.-W. Huang, and J. Li,
Nat. Photonics \textbf{6}, 866 (2012).

\end{thebibliography}
\end{document}